\documentclass[doublecol]{epl2} 

\title{$\mathcal{PT}$-symmetric quantum oscillator in an optical cavity}
\shorttitle{$\mathcal{PT}$-symmetric quantum oscillator  ... } 

\author{S. Longhi \inst{1,2}}
\shortauthor{S. Longhi}

\institute{                    
  \inst{1}  Dipartimento di Fisica, Politecnico di Milano, Piazza L. da Vinci 32, I-20133 Milano, Italy\\
  \inst{2}  Istituto di Fotonica e Nanotecnlogie del Consiglio Nazionale delle Ricerche, sezione di Milano, Piazza L. da Vinci 32, I-20133 Milano, Italy}

\abstract{The quantum harmonic oscillator with parity-time ($\mathcal{PT}$) symmetry, obtained from the ordinary (Hermitian) quantum harmonic oscillator by an imaginary displacement of the spatial coordinate, provides an important and exactly-solvable model to investigate non-Hermitian extension of the Ehrenfest theorem. Here it is shown that transverse light dynamics in an optical resonator with off-axis longitudinal pumping can emulate a $\mathcal{PT}$-symmetric quantum harmonic oscillator, providing an experimentally accessible system to investigate non-Hermitian coherent state propagation.}
\begin{document}

\maketitle

\section{Introduction}

The concept of parity-time ($\mathcal{PT}$) symmetry in optics has received a rapidly growing attention in the past decade since the publication of a few seminal papers \cite{r1,r2,r3}, opening up interesting applications and prospects in photonics (see, for instance, \cite{r4,r5,r6,r7,r8,r9,r10,r10bis,r11,r12,r13,r14,r15,r16} and references therein). As a matter of fact, the recent experimental progress on $\mathcal{PT}$ symmetry in optics is considered one of the most important achievements in recent physical research \cite{r17,r18}.\\ 
Optical structures with balanced gain and loss regions provide a fertile laboratory tool to visualize important physical implications of
$\mathcal{PT}$ symmetry and non-Hermitian dynamics which do not have a counterpart in conservative (Hamiltonian) dynamics. For example, the classical limit of the $\mathcal{PT}$-symmetric quantum dynamics, described by generalized canonical equations, shows an interesting geometric structure \cite{r19,r20} which provides a nontrivial extension of the Ehrenfest theorem. As shown in a series of papers by E.M. Graefe {\it et al.} \cite{r19,r20,r21,r22}, the generalized canonical equations involve a metric gradient flow, in addition to the usual Hamiltonian (conservative) dynamics. The gradient flow is coupled to an evolution equation for the metric, which in turn depends on the phase space coordinates. Such a geometrical structure of coherent state dynamics has been studied in details for a few important models, including the dissipative anharmonic oscillator \cite{r20}, the non-Hermitian Swanson oscillator \cite{r22}, and the driven Hatano-Nelson lattice displaying Bloch oscillations in the complex plane \cite{r23,r24}.\\  
In this Letter we consider the semiclassical dynamics of a simple and exactly-solvable $\mathcal{PT}$-symmetric model, namely a $\mathcal{PT}$-symmetric quantum harmonic oscillator (QHO), which is obtained from the ordinary (Hermitian) QHO by an imaginary displacement of the spatial coordinate \cite{r25,r26,r27}. In this model the signature of non-Hermitian classical dynamics is the appearance of an {\it anharmonic} motion of Gaussian (coherent) states. Here we suggest to emulate the $\mathcal{PT}$-symmetric QHO and to visualize the anharmonic motion of coherent states by considering transverse light dynamics in an optical cavity. Optical resonators have been deeply investigated in the context of laser science \cite{r28}, and non-Hermitian signatures in such systems have been mostly focused to the laser linewidth enhancement factor in certain cavities with non-orthogonal modes (see, for instance, \cite{r40,r41,r42,r43} and references therein). Recent works have suggested and experimentally demonstrated that beam dynamics in optical cavities and lensguides can effectively simulate the quantum mechanical Schr\"{o}dinger equation in rather extended forms \cite{r29,r30,r31,r32,r33,r34,r35}, for example, to emulate a fractional kinetic energy
operator \cite{r30,r31,r34}, synthetic magnetic fields \cite{r32,r33}, and quantum dissipation \cite{r35}. Here we show that optical cavities can provide a fertile laboratory tool to investigate the classical limit of $\mathcal{PT}$-symmetric quantum dynamics as well.

\section{$\mathcal{PT}$-symmetric quantum harmonic oscillator and coherent state dynamics}  
The Hamiltonian of the $\mathcal{PT}$-symmetric QHO is obtained from the ordinary (Hermitian) QHO after an imaginary displacement of the spatial coordinate, and reads \cite{r25,r26,r27}
\begin{equation}
\hat{H}=-\frac{\hbar^2}{2m} \frac{d^2}{dx^2}+ \frac{1}{2} m \Omega^2 (x-i \delta)^2
\end{equation}
where $m$ and $\Omega$ are the mass and frequency of the oscillator, and $i \delta$ (with $\delta$ real) is the imaginary displacement of position. The Hamiltonian (1) is always in the unbroken $\mathcal{PT}$ phase, since it is isospectral to the Hermitian QHO, with eigenfunctions given by the usual Gauss-Hermite modes with complex argument $x-i\delta$.   
 The generalized canonical equations for the mean values of particle position $q$ and momentum $p$, coupled to the covariance matrix $\Sigma$, can be written in terms of the Hermitian and anti-Hermitian parts of $\hat{H}$ \cite{r20,r22,r24}. For the $\mathcal{PT}$-symmetric QHO they read explicitly  
\begin{eqnarray}
\frac{dq}{dt} =  \frac{p}{m}-m \Omega^2 \delta \Sigma_{qq} \; , \;  
\frac{dp}{dt} =  - m \Omega^2 q- m \Omega^2 \delta  \Sigma_{pq}\\
\frac{d \Sigma_{qq}}{dt}  =  \frac{2}{m} \Sigma_{pq} , \;\;  \frac{d \Sigma_{pp}}{dt}  = -2 m \Omega^2  \Sigma_{pq} \\
 \frac{d \Sigma_{pq}}{dt}  =  \frac{1}{m} \Sigma_{pp}-m \Omega^2 \Sigma_{qq}.
\end{eqnarray}
It should be noted that, since the Hamiltonian (1) is quadratic in $x$ and $-i \partial_x$, the classical dynamics, governed by the generalized canonical equations (2-4),  exactly describes the quantum dynamics of Gaussian wave packets \cite{r22}. Indeed, the exact propagator of $\hat{H}$ is simply given by the propagator of the Hermitian QHO (the Mehler kernel) after complex spatial displacement of spatial coordinates. The general solution to the Schr\"{o}dinger equation $i \hbar \partial_t \psi(x,t) = \hat{H} \psi$ is hence given by
\begin{equation}
\psi(x,t)= \int_{-\infty}^{\infty} d \xi U(x-i \delta, \xi-i \delta, t) \psi( \xi, t)
\end{equation} 
where $U(x, \xi, t)$ is the Mehler kernel
\begin{eqnarray}
U(x,\xi,t) =  \sqrt{\frac{m \Omega}{2 \pi i \hbar \sin (\Omega t)}} \\
 \times  \exp \left\{  \frac{i m \Omega}{2 \hbar \sin (\Omega t) } \left[  (x^2 + \xi^2) \cos (\Omega t) -2 x \xi \right] \right\}. \nonumber
\end{eqnarray}
Let us assume as an initial condition a Gaussian wave packet (coherent state), namely
\begin{equation}
\psi(x,0) \propto \exp [-\sigma (x-q_0)^2+i p_0x / \hbar]
\end{equation}
with center of mass $q_0$, initial momentum $p_0$ and localization length $ \sim 1 / \sqrt{\sigma}$. Substitution of Eqs.(6) and (7) into Eq.(5) and after performing the integral, a Gaussian wave packet for $\psi(x,t)$ with time-varying parameters is obtained. In particular, the center of mass $q(t)$ of the Gaussian wave packet can be written as $q(t)=q^{(H)}(t)+q^{(NH)}(t)$,
where we have set 
\begin{eqnarray}
q^{(H)}(t)=q_0 \cos (\Omega t)+\frac{p_0}{m \Omega} \sin (\Omega t) \\
q^{(NH)}(t)=\delta \left(\frac{\sigma \hbar}{m \Omega}-\frac{m \Omega}{4 \sigma \hbar} \right)\sin(2 \Omega t) -\frac{2 \delta \sigma \hbar}{m \Omega} \sin (\Omega t).
\end{eqnarray}
The same result could have been obtained by solving the generalized canonical equations (2-4). 
Note that $q^{(H)}(t)$ is the usual sinusoidal path of a coherent state in the Hermitian QHO, i.e. for $\delta =0$, whereas $q^{(NH)}(t)$ is a correction that accounts for the anti-Hermitian part of $\hat{H}$. Interestingly, the non-Hermitian correction $q^{(NH)}(t)$ to the classical trajectory contains a term oscillating {\it twice} the QHO frequency $\Omega$, thus making the resulting path {\it anharmonic}. Anharmonic motion of a coherent state in the $\mathcal{PT}$-symmetric QHO can be thus regarded as a clear signature of non-Hermitian classical dynamics. 

\begin{figure}
\onefigure[width=8cm]{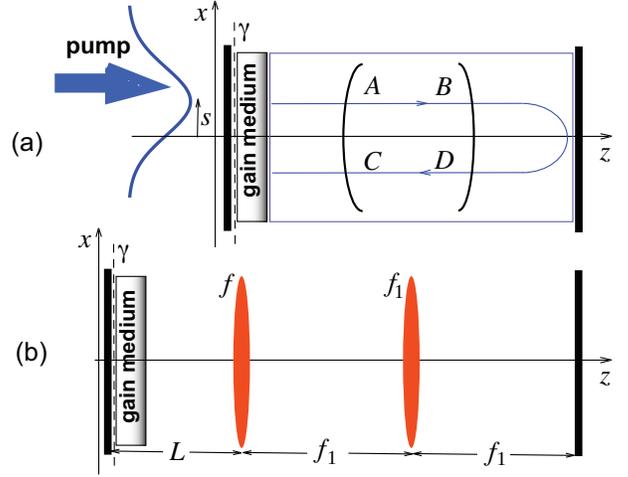}
\caption{(Color online) (a) Schematic of a generic Fabry-Perot optical resonator with off-axis longitudinal pumping that realizes the $\mathcal{PT}$-symmetric quantum harmonic oscillator. $ABCD$ is the resonator round-trip matrix with respect to the plane $\gamma$, with $A=D$. (b) Optical resonator used to simulate non-Hermitian anharmonic motion of a Gaussian beam. The resonator is stable for $L<f$.}
\end{figure}

\section{Optical realization of the $\mathcal{PT}$-symmetric quantum oscillator} 
Here we propose an optical realization of the $\mathcal{PT}$-symmetric QHO [Eq.(1)], which is based on transverse light dynamics in an optical resonator filled with an active medium with spatially-inhomogeneous gain. A schematic of the Fabry-Perot optical cavity is shown in Fig.1(a). For the sake of simplicity, we will consider one transverse spatial coordinate $x$, however the analysis can be extended {\it mutatis mutandis} to the two-dimensional case. The optical cavity comprises a flat end mirror (transmittance $T$) on the left hand side, a thin gain medium placed close to the flat mirror, and a generic geometrically-stable sequence of spherical optical elements, which is described by a round-trip matrix $ABCD$ with respect to the plane $\gamma$ [see Fig.1(a)]. Since the optical sequence is crossed by an optical ray twice in opposite directions, one has $A=D$. Stability of the cold cavity is ensured provided that the angle $\theta$, defined by the relation $\cos \theta=(A+D)/2=A$, is real, i.e. for $|A|<1$ \cite{r28}.  
 The gain medium provides a spatially-dependent gain coefficient $g(x)$ in each round-trip, which is determined by the pumping configuration. For example, in a longitudinally-pumped solid-state laser \cite{iff} $g(x)$ is determined by the Gaussian distribution of the pump beam [Fig.1(a)]. We also assume in the most general case that an optical pulsed field $\psi^{(in)}(x,t)=A(t)F(x)$, with pulse envelope $A(t)$ and transverse spatial distribution $F(x)$, is externally injected into the optical cavity through the flat end mirror. 
 Indicating by $l$ the round-trip logarithmic loss of the cavity and by $\lambda=2 \pi / k$ the wavelength of light, the evolution of the transverse beam distribution $\psi_n(x)$ at successive transits in the resonator is governed by the following map \cite{r28,r30,r31,r35} 
 \begin{eqnarray}
 \psi_{n+1}(x) & = & \exp(-l) \exp[g(x)/2] \hat{K} \exp[g(x)/2]  \psi_n(x) \nonumber \\
 & + & \sqrt{T} A_n F(x) \exp(i n \Delta) \
 \end{eqnarray} 
where $n$ is the round-trip number, $\psi_n(x)$ is the field distribution at the reference plane $\gamma$, $A_n=A(nT_R)$, $T_R$ is the transit time of light in the resonator, $\Delta$ is a detuning parameter that accounts for the offset between the carrier frequency of the injected field and the axial frequency of the cavity,  and $\hat{K}$ is the Fresnel integral operator that describes beam propagation in the $ABCD$ optical sequence, i.e. $\hat{K} f(x)=\int d \xi \mathcal{K}(x, \xi) f( \xi)$ with the Fresnel kernel \cite{r28}
\begin{equation}
\mathcal{K}(x, \xi)= \sqrt{\frac{i}{\lambda B}} \exp \left[ - i\frac{\pi}{\lambda B} \left( Dx^2+A \xi^2 -2 \xi x \right) \right]
\end{equation}
Let us assume that the gain profile $g(x)$ varies slowly around the optical axis $x=0$ as compared to the characteristic spot size $w_0$ of the fundamental TEM$_{\rm 00}$ Gaussian  resonator mode. In this case we can approximate $g(x)$ up to first order around $x=0$ as $g(x) \simeq g_0+ \alpha x$, where $g_0 \equiv g(0)$ and $\alpha \equiv (dg/dx)_{x=0}$. We will assume $\alpha \neq 0$. For example, in a longitudinal pumping configuration typical of solid-state lasers \cite{iff} with the pump beam displaced by $s$ from the optical axis of the cavity [see Fig.1(a)], one has $g(x)=g_p \exp[-2(x-s)^2/w_p^2]$, where $w_p \gg w_0$ is the pump beam spot size and $g_p$ the peak gain. Assuming an off-axis displacement $s=w_p/2$, so as $(d^2g/dx^2)_{x=0}=0$ and the approximation is valid up to second order, one has $g_0=g_p / \sqrt{e}$ and 
\begin{equation}
\alpha=\frac{2 g_p}{\sqrt{e} w_p}=\frac{2 g_0}{w_p}.
\end{equation}   
To show the equivalence between transverse light dynamics in the optical cavity and the $\mathcal{PT}$-symmetric QHO, let us consider the limiting case of a vanishing injected field ($A_n=0$), so that after setting $g(x) \simeq g_0+ \alpha x$ Eq.(10) takes the form 
\begin{equation}
\psi_{n+1}(x)=\exp(g_0-\gamma) \hat{L} \psi_n(x),
\end{equation}
 where we have set $\hat{L} \equiv \exp(\alpha x/2) \hat{K} \exp(\alpha x/2)$. For our purposes, it is worth writing Eq.(13) as a stroboscopic map for an associated Schr\"{o}dinger-like wave equation. To this aim, we write the operator $\hat{L}$ in the exponential form $\hat{L}= \exp(-i k \hat{H})$, so that $\psi_n(x)$ is obtained as $\psi_n(x)=\psi(x,t=n)$, where $\psi(x,t)$ satisfies the Schr\"{o}dinger-like wave equation 
 \begin{equation}
 \frac{i}{k} \frac{\partial  \psi}{\partial t}=\hat{H} \psi+\frac{i}{k}(g_0-l) \psi. 
\end{equation}
The derivation of the explicit form of the operator $\hat{H}$ is rather cumbersome and can be done following a procedure similar to the one presented in Ref.\cite{r36}. One obtains
\begin{equation}
\hat{H}= \frac{\theta B}{2k^2 \sin \theta} \frac{d^2}{dx^2}-\frac{1}{2} \frac{\sin \theta}{ \theta B} \theta^2 (x-i \delta)^2,
\end{equation}
where we have set $\delta \equiv - \alpha (1+A)/(2 k C)$ and $\cos \theta=A$ \footnote{In writing Eq.(15), we disregarded an additional constant and real energy term, that just provides an energy shift to the spectrum of $\hat{H}$, i.e. a shift of the resonance frequencies of the optical cavity.}. Clearly, for $g_0=l$ Eq.(14) emulates the $\mathcal{PT}$-symmetric QHO [Eq.(1)] with frequency $\Omega$ and mass $m$ given by
\begin{equation}
\Omega=\theta \; , \;\;\;  m=-\sin \theta /( \theta B),
\end{equation}
provided that the formal substitution $\hbar \rightarrow 1/k= \lambda / (2 \pi)$ is made. Note that the Hermitian limit $\delta=0$ of the QHO is just obtained in our optical setting by assuming $s=0$, i.e. by on-axis pumping. Note also that, since $\hat{H}$ is always in the unbroken $\mathcal{PT}$ phase, Eq.(14) shows that the laser threshold is attained at $g_{0 \; th}=l$.

\begin{figure}
\onefigure[width=8cm]{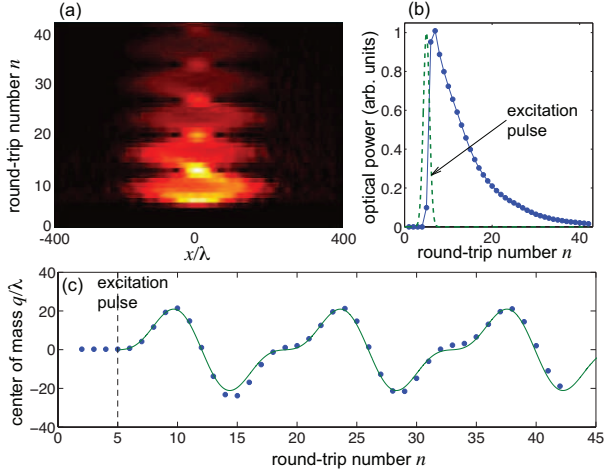}
\caption{(Color online) Numerically-computed transverse beam evolution in the resonator of Fig.1(b) after excitation with a short Gaussian pulsed beam. (a) Evolution of the field amplitude $|\psi_n(x)|$ (pseudo color map) at successive round trips $n$ in the resonator. (b) Evolution of the optical beam power $P=\int dx |\psi_n(x)|^2$ (circles) versus $n$ during and after pulsed beam excitation. The dashed curve shows the temporal envelope $A(t/T_R)$ of the excitation beam. Since the optical cavity is operated below lasing threshold, $P$ decays after external pulse excitation. (c) Behavior of the beam center of mass $q(n)$, in units of light wavelength $\lambda$, versus round-trip number $n$ (dots) and theoretical curve as predicted by Eqs.(8) and (9) (solid curve). The vertical dashed line shows the instant of external pulse excitation. }
\end{figure}

\section{Signatures of non-Hermitian dynamics}
Having established an analogy between the $\mathcal{PT}$-symmetric QHO and light dynamics in the optical cavity of Fig.1(a), let us discuss some signatures of non-Hermitian light dynamics.\\
{\it Anharmonic Gaussian motion.} The anharmonic trajectory of a coherent state in the  $\mathcal{PT}$-symmetric QHO is the main signature of non-Hermitian classical dynamics, as discussed above. Such an effect can be visualized by considering light dynamics at successive transits in the resonator, when the the optical cavity is kept below laser threshold ($g_0<l$) and it is excited by a pulsed Gaussian beam $\psi^{(in)}(x,t)=A(t) \exp(- x^2/w_e^2)$, with spot size $w_e \neq w_0$ and pulse duration shorter than the cavity round-trip time. The free-decay of light in the cavity following the short pulse excitation basically emulates the evolution of the $\mathcal{PT}$-symmetric QHO with the initial condition $\psi(x,0)=\exp(-x^2 / w_e^2)$. In an experiment, transverse light evolution at successive transits in the cavity can be detected using a gated camera as demonstrated e.g. in Refs.\cite{r37,r38,r39}. As an example, let us consider the optical cavity shown in Fig.1(b), comprising two flat end mirrors and two focusing lenses of focal lengths $f$ and $f_1$. The elements of the round-trip matrix of the resonator are $A=D=2L/f-1$, $B=2L(L/f-1)$ and $C=2/f$, indicating that the resonator is stable for $L<f$. Note that the stability and light dynamics in the resonator is independent of $f_1$, however the use of the second lens $f_1$ in Fourier configuration increases the transit time $T_R$ of light in the resonator (e.g. from few to few tens of ns), which could be helpful in an experiment to get time-resolved transverse beam frames. Figure 2 shows the excitation and subsequent decay of the intracavity light field at plane $\gamma$ as obtained by numerical solution of the map (10) for an off-axis Gaussian pumping $g(x)=g_p \exp[-2(x-s)^2/w_p^2]$ for parameter values $f/ \lambda=10^5$, $L/f=0.95$, $l=0.18$, $g_p=0.2$, $w_p/ \lambda=483$, and $s=w_p/2$, corresponding to $g_0 \simeq 0.12 $ (the laser is below threshold). The excitation field $\psi^{(in)}(x,t)$ is a pulsed Gaussian beam $\psi^{(in)}(x,t)=A(t) \exp(-x^2/w_e^2)$ with spot size $w_e= 40 \lambda$ and Gaussian pulse envelope $A(t)=\exp[-(t-t_p)^2/ \tau_p^2]$  with $t_p=5T_R$ and pulse duration $\tau_p=T_R$. For such parameter values, the frequency of the harmonic motion is $\Omega= \theta={\rm arcos}(A) \simeq 0.45$, corresponding to a classical period $T_{HO}= 2 \pi / \Omega \simeq 14$ round trips, the oscillator mass is $m=-\sin \theta / (B \theta) \simeq (1.017 / \lambda)  \times 10^{-4}$, whereas the spatial displacement $\delta$ turns out to be $\delta =-\alpha(1+A)/(2kC)=-g_p L \lambda /(2 \pi \sqrt{e} w_p) \simeq -3.8 \lambda$. Figure 2(a) shows the evolution of the beam amplitude $|\psi_n(x)|$ at successive transits in the resonator, whereas Fig.2(b) depicts the evolution of the intracavity power $P(n)=\int dx |\psi_n(x)|^2$ versus $n$ (circles) and of the excitation pulse envelope $A(t)$ (dashed curve). The motion of the beam center of mass $q=q(n)$ is shown in Fig.2(c) by circles and compared to the theoretical prediction $q^{(H)}(t)+q^{(NH)}(t)$ of the $\mathcal{PT}$ QHO [Eqs.(8) and (9)]. The figure clearly indicates that the Gaussian motion is anharmonic as a signature of the non-Hermitian complex displacement $\delta$, and that the discretized dynamics at the successive round-trips in the resonator system emulates with an excellent accuracy the $\mathcal{PT}$-symmetric  QHO dynamics.\\
\\
{\it Tilted laser emission.} An interesting property of the fundamental (lowest-energy) Gaussian eigenmode of the $\mathcal{PT}$-symmetric QHO is that there is a continuous energy flow from the 'gain' to the 'lossy' regions, which results is a non-vanishing mean momentum despite the state is stationary in time \cite{r27}. In fact, the fundamental Gaussian mode of the $\mathcal{PT}$-symmetric QHO is simply given by
\begin{equation}
\psi_0(x) \propto \exp[-\rho(x-i \delta)^2] \propto \exp(- \rho x^2+2i \rho \delta x) 
\end{equation}
 which possesses a nonvanishing momentum $ p_0=2 \hbar \rho \delta= m \Omega \delta$, where $\rho \equiv m \Omega / 2 \hbar$. Let us now suppose that at some time $t=t_0$ the trapping (parabolic) potential is suddenly switched off, so that for $t>t_0$ the Gaussian state is a freely-evolving wave packet. Owing to the nonvanishing mean momentum $p_0$, the wave packet at $t>t_0$ is not anymore at rest and moves with a constant speed $v_0=p_0/m=\Omega \delta $.  In our optical realization of the  $\mathcal{PT}$ QHO, such an interesting effect can be simply detected by observing the laser beam that escapes from the optical cavity when it is lasing in the fundamental TEM$_{\rm 00}$ transverse mode. The transverse motion of the emitted beam indicates that the laser emission is tilted with respect to the axis $z$ of the cavity by the angle
 \begin{equation}
 \theta_{tilt}=\frac{\lambda \delta}{\pi w_0^2}
 \end{equation}
 where $w_0=1/ \sqrt{\rho}=\sqrt{|2B/( k \sin \theta|)}$ is the beam waist of the TEM$_{\rm 00}$ Gaussian mode at plane $\gamma$. It is worth comparing the tit angle $\theta_{tilt}$ with the divergence angle $\theta_{d}= \lambda / (\pi w_0)$ of the Gaussian beam. One obtains
 \begin{equation}
 \frac{\theta_{tilt}}{\theta_d}= \frac{\delta}{w_0}.
 \end{equation}
 In a stable resonator and for typical parameter values, $\delta$ turns out to be smaller (or much smaller) than $w_0$, so that the tilt angle induced by the spatial displacement $\delta$ is smaller than the divergence angle. For example, for the resonator of Fig.1(b) and parameter values used in the simulations of Fig.2, one has $\theta_{tilt} / \theta_d \simeq 0.075$. Nevertheless, such a small tilt of laser emission as the pumping beam is displaced from the optical axis should be detected in far-field measurements.\\
  \\
 {\it Non-Hermitian enhancement of laser linewidth.} 
 The eigenfunctions of the $\mathcal{PT}$ QHO, given by the ordinary Gauss-Hermite modes of the QHO with complex argument, do not form a set of orthogonal modes. In laser physics, an important consequence of the lack of orthogonality of laser modes is an enhancement of the Schawlow-Townes laser linewidth, which is expressed by the so-called Petermann excess noise factor $K$ (see, for instance, \cite{r40,r41,r42,r43} and references therein). For the laser oscillating in the fundamental transverse mode, the excess noise factor is given by $K=\langle \psi_0, \psi_0 \rangle   \langle \psi^{\dag}_0, \psi^{\dag}_0 \rangle /  | \langle \psi_0, \psi^{\dag}_0 \rangle|^2$, where $\psi_0(x)$ is given by Eq.(17) and $\psi^{\dag}_0(x)$ is its Hermitian adjoint eigenmode, which is simply obtained from Eq.(17) by the change $\delta \rightarrow -\delta$. One readily obtains $K=\exp[(2 \delta / w_0)^2]=\exp[4 (\theta_{tilt} / \theta_{d})^2]$. For typical resonator parameters that realize the $\mathcal{PT}$-symmetric QHO, $\theta_{tilt} / \theta_{d}$ is smaller that one as discussed above, and the excess noise factor turns out to be modest or even negligible ($K \simeq 1.023$ for the example discussed above). Therefore, unlike lasers with unstable resonators leading to large excess noise factors \cite{r40,r41,r42}, in our optical setting non-Hermitian enhancement of the laser linewidth is not a major effect.

  \section{Conclusion} The  $\mathcal{PT}$-symmetric quantum harmonic oscillator provides an important and exactly-solvable model to investigate non-Hermitian extension of the Ehrenfest theorem. The main signature of non-Hermitian dynamics here is provided by the {\it anharmonic} motion of a coherent state. 
In this work we have theoretically  shown that transverse dynamics of light waves in longitudinally-pumped optical cavities, such as those generally used in end-pumped solid-state lasers,  realizes in optics a $\mathcal{PT}$-symmetric quantum harmonic oscillator, paving the way toward an experimental observation of non-Hermitian coherent state propagation.


\end{document}